\documentclass[%
 reprint,
superscriptaddress,
nobibnotes,
 aps,
pre,
floatfix,
]{revtex4-1}

\usepackage{amsmath,amssymb}
\usepackage{dsfont}
\usepackage{stackengine,graphicx}
\usepackage{dcolumn}
\usepackage{bm}
\usepackage{bbm}
\usepackage{soul}
\usepackage{breqn}
\usepackage{subfigure}
\usepackage{float}
\usepackage{flushend}
\usepackage{balance}
\usepackage[dvipsnames]{xcolor} 

\usepackage{hyperref}
\usepackage{cleveref}

\crefname{equation}{Eq.}{Eqs.}
\crefname{figure}{Fig.}{Figs.}

\newcommand{\be}{\begin{equation}}
\newcommand{\ee}{\end{equation} }

\RequirePackage[normalem]{ulem} 
\RequirePackage{color}\definecolor{RED}{rgb}{1,0,0}\definecolor{BLUE}{rgb}{0,0,1} 
\providecommand{\DIFaddbegin}{} 
\providecommand{\DIFaddend}{} 
\providecommand{\DIFdelbegin}{} 
\providecommand{\DIFdelend}{} 
\providecommand{\DIFaddbeginFL}{} 
\providecommand{\DIFaddendFL}{} 
\providecommand{\DIFdelbeginFL}{} 
\providecommand{\DIFdelendFL}{} 
\newcommand{\DIFscaledelfig}{0.5}
\RequirePackage{settobox} 
\RequirePackage{letltxmacro} 
\newsavebox{\DIFdelgraphicsbox} 
\newlength{\DIFdelgraphicswidth} 
\newlength{\DIFdelgraphicsheight} 
\LetLtxMacro{\DIFOincludegraphics}{\includegraphics} 
\newcommand{\DIFaddincludegraphics}[2][]{{\color{blue}\fbox{\DIFOincludegraphics[#1]{#2}}}} 
\newcommand{\DIFdelincludegraphics}[2][]{
\sbox{\DIFdelgraphicsbox}{\DIFOincludegraphics[#1]{#2}}
\settoboxwidth{\DIFdelgraphicswidth}{\DIFdelgraphicsbox} 
\settoboxtotalheight{\DIFdelgraphicsheight}{\DIFdelgraphicsbox} 
\scalebox{\DIFscaledelfig}{
\parbox[b]{\DIFdelgraphicswidth}{\usebox{\DIFdelgraphicsbox}\\[-\baselineskip] \rule{\DIFdelgraphicswidth}{0em}}\llap{\resizebox{\DIFdelgraphicswidth}{\DIFdelgraphicsheight}{
\setlength{\unitlength}{\DIFdelgraphicswidth}
\begin{picture}(1,1)
\thicklines\linethickness{2pt} 
{\color[rgb]{1,0,0}\put(0,0){\framebox(1,1){}}}
{\color[rgb]{1,0,0}\put(0,0){\line( 1,1){1}}}
{\color[rgb]{1,0,0}\put(0,1){\line(1,-1){1}}}
\end{picture}
}\hspace*{3pt}}} 
} 
\LetLtxMacro{\DIFOaddbegin}{\DIFaddbegin} 
\LetLtxMacro{\DIFOaddend}{\DIFaddend} 
\LetLtxMacro{\DIFOdelbegin}{\DIFdelbegin} 
\LetLtxMacro{\DIFOdelend}{\DIFdelend} 
\DeclareRobustCommand{\DIFaddbegin}{\DIFOaddbegin \let\includegraphics\DIFaddincludegraphics} 
\DeclareRobustCommand{\DIFaddend}{\DIFOaddend \let\includegraphics\DIFOincludegraphics} 
\DeclareRobustCommand{\DIFdelbegin}{\DIFOdelbegin \let\includegraphics\DIFdelincludegraphics} 
\DeclareRobustCommand{\DIFdelend}{\DIFOaddend \let\includegraphics\DIFOincludegraphics} 
\LetLtxMacro{\DIFOaddbeginFL}{\DIFaddbeginFL} 
\LetLtxMacro{\DIFOaddendFL}{\DIFaddendFL} 
\LetLtxMacro{\DIFOdelbeginFL}{\DIFdelbeginFL} 
\LetLtxMacro{\DIFOdelendFL}{\DIFdelendFL} 
\DeclareRobustCommand{\DIFaddbeginFL}{\DIFOaddbeginFL \let\includegraphics\DIFaddincludegraphics} 
\DeclareRobustCommand{\DIFaddendFL}{\DIFOaddendFL \let\includegraphics\DIFOincludegraphics} 
\DeclareRobustCommand{\DIFdelbeginFL}{\DIFOdelbeginFL \let\includegraphics\DIFdelincludegraphics} 
\DeclareRobustCommand{\DIFdelendFL}{\DIFOaddendFL \let\includegraphics\DIFOincludegraphics} 

\begin{document}

 \title{Inverse problems for structured datasets using parallel TAP equations and RBM}

\makeatletter
\let\cat@comma@active\@empty
\makeatother

\author{Aurelien Decelle}\DIFaddbegin \affiliation{Laboratoire de Recherche en Informatique, TAU - INRIA, CNRS,  Universit\'e Paris-Sud et Universit\'e Paris-Saclay, B\^at. 660, 91190 Gif-sur-Yvette, France}
\DIFaddend 

\author{Sungmin Hwang}

\author{Jacopo Rocchi}
\affiliation{LPTMS, Universit\`{e} Paris-Sud 11, UMR 8626 CNRS, B\^{a}t. 100, 91405 Orsay Cedex, France}

\author{Daniele Tantari}
\affiliation{University of Florence, department of economics and management}


\date{\today}

\begin{abstract}

We propose an efficient algorithm to solve inverse problems in the presence of binary clustered datasets.
We consider the paradigmatic Hopfield model in a teacher student scenario, where this situation is found in the retrieval phase.
This problem has been widely analyzed through various methods such as mean-field approaches or the pseudo-likelihood optimization.
Our approach is based on the estimation of the posterior using the Thouless-Anderson-Palmer (TAP) equations in a parallel updating scheme. At the difference with other methods, it allows to retrieve the exact patterns of the teacher and the parallel update makes it possible to apply it for large system sizes.
We also observe that the Approximate Message Passing (AMP) equations do not reproduce the expected behavior in the direct problem, questioning the standard practice used to obtain time indexes coming from Belief Propagation (BP).
We tackle the same problem using a Restricted Boltzmann Machine (RBM) and discuss the analogies and the differences between our algorithm and the RBM learning. 

\end{abstract}

\pacs{Valid PACS appear here}
\maketitle

Inverse problems consist in inferring information about the structure of a system from the observation data of configurations.
Cases where the system's variables $s_i$ are binary can be studied in the framework of the inverse Ising model, whose parameters $\{J_{ij},h_i\}$ are tuned in order to describe the observed configurations according to the Boltzmann weight $P(s) \sim \exp  (\sum_{i<j} J_{ij} s_i s_j + \sum_i h_i s_i)$. 
This is the simplest distribution emerging when using the maximum entropy approach in order to reproduce exactly the  one and two points statistics of the data.
One of the most successfull application of this method is the 3D reconstruction of protein structures \cite{morcos2011direct}. Inverse problems arise also in collective behavior \cite{bialek2014social}, immunology \cite{ wood2012mechanism}, neural activity \cite{schneidman2006weak, cocco2009neuronal} and financial time series \cite{bury2013market}.
In general, inferring the parameters of the model is a challenging problem because maximizing the likelihood involves the computation of the partition function $Z=\sum_s P(s)$, which is impossible in most of the realistic cases.
On the other hand, when dealing with time-series, it is possible to use a simples approach modeling the sampling process by a stochastic parallel dynamics analysed in \cite{roudi2011mean}, optimized in \cite{decelle2015inference} and generalized in \cite{dunn2013learning, campajola2018inference}. A recent review on this subject can be found in \cite{nguyen2017inverse}.

The original attempt to solve the problem was based on an Expectation-Maximization algorithm known as Boltzmann learning \cite{ackley1985learning}. This method is unpractical on large systems unless heuristic methods, like Monte Carlo (MC) sampling, are used to estimate correlations \cite{huang2010reconstructing}.
Nevertheless MC is slow and thus many sophisticated techniques coming from statistical mechanics and machine learning have been proposed as alternative approaches \cite{kappen1998efficient,tanaka2000information,sohl2011new,cocco2011adaptive,aurell2012inverse,ricci2012bethe,nguyen2012mean,cocco2012adaptive,raymond2013mean,
decelle2014pseudolikelihood,lokhov2018optimal,franz2019fast}.
These methods, however, share one or both of the following shortcomings: i) they require a large number of observations and ii) the overall performance drops significantly when the dataset is structured. This is often the case when data is produced from a (sub)set of many attractive states, or data is collected in different regimes, e.g. quiescent and spiking regimes in neural networks.
This problem becomes particularly relevant at low temperatures and it has already been studied both in the sparse \cite{braunstein2011inference} and in the dense case \cite{cocco2011high,decelle2016solving}. 
Pseudo-likelihood \cite{besag1977efficiency} based methods \cite{decelle2016solving} were shown to be the best options in a wide range of temperatures.
Here, we present two algorithms to compete with the existing state-of-the-art by
posing the problem in a Bayesian framework using the Thouless-Anderson-Palmer (TAP) equations \cite{thouless1977solution} and the Restricted Boltzmann Machine (RBM) \cite{hinton2002training}.
Our TAP-based algorithm will be shown to achieve a better quality of the results by observing far fewer configurations in the clusterized phase. Moreover, it allows to consider larger system size with respect to those studied in \cite{cocco2011high,decelle2016solving}.
As a side aspect of this work, we discuss the behavior of the Approximate Message Passing (AMP) equations for the Hopfield model \cite{mezard2017mean}, and show that they do not do not reproduce the thermodynamic expected behavior.

We consider  a dataset with many clusters by drawing configurations from  the Hopfield model \cite{hopfield1982neural, amit1985storing}. Given a set of $N$-dimensional binary independent patterns  $\{\underline{\zeta}^{\mu}\}_{\mu=1,\dots,P}$, teacher's patterns, the coupling matrix of the associated Hopfield model is defined as $J_{ij} = N^{-1} \sum_{\mu} \zeta^{\mu}_i \zeta^{\mu}_j $ and its Hamiltonian is $H_{\zeta}(\underline{s})=-1/2 \sum_{ij} J_{ij} s_i s_j $. 
Next, we construct a set of equilibrium configurations $\mathcal{D} = \{\underline{s}^a\}_{a=1,\dots,M}$ sampled from the Boltzmann distribution $P(\underline{s}) = Z^{-1}\exp[-\beta H_{\zeta}(\underline{s})]$, $\beta$ being the inverse temperature.
The set $\mathcal{D}$ is given to a student whose task is to infer the teacher's patterns. This task cannot be accomplished with the methods developed in \cite{cocco2011high,decelle2016solving}, where the focus was the inference of the coupling matrix $J$. 
Using a uniform prior on teacher's patterns, the posterior distribution is proportional to the likelihood.

For $P=1$, the Hopfield model is nothing but a Curie-Weiss model. 
In this case, the problem is called the \textit{dual} Hopfield model \cite{barra2017phase, barra2018phase}, and the log-likelihood is $\mathcal{L}(\underline{\xi}|\mathcal{D}) = \frac{\beta}{2N}  \sum_{ij} \sum_{a=1}^M s^a_i s^a_j \xi_i \xi_j \:$, where  $\underline{\xi}$ denote the student's pattern.
This is readily established by absorbing the $\xi$-dependence of the partition function into a redefined set of variables $\underline{s}$ via $s'_i=\xi_i s_i$.
On the other hand, for $P>1$ this transformation is not feasible. In this case, the log-likelihood reads
\begin{dmath}
	\mathcal{L} (\{\underline{\xi}^{\mu}\}_{\mu=1,\dots,P}|\mathcal{D}) =  \frac{\beta}{2N}  \sum_{ij} \sum_{\mu=1}^P \sum_{a=1}^M s^a_i s^a_j \xi^{\mu}_i \xi^{\mu}_j   -M \log Z (\{\xi^{\mu}_i\}_{\mu=1,\dots,P}) \:.
	\label{eq:dualHMP}
\end{dmath}

TAP equations \cite{mezard1990spin,nakanishi1997mean,shamir2000thouless,kabashima2001tap,mezard2017mean} describe the stationary points of the free energy and their use as an inference method has been pioneered in \cite{opper1996mean, kappen1998boltzmann, tanaka1998mean}, as well as in \cite{kabashima1998belief, kabashima2001tap}, where their relationship between the Bayes theorem and the Belief Propagation (BP) equations was discussed. These works paved the way to their applications in a number of other problems \cite{saad2009line} such as error correcting codes, compressed sensing and learning in neural networks, as discussed in the recent review \cite{zdeborova2016statistical}. 

\begin{figure}[h]
\centering
    \includegraphics[scale=0.65]{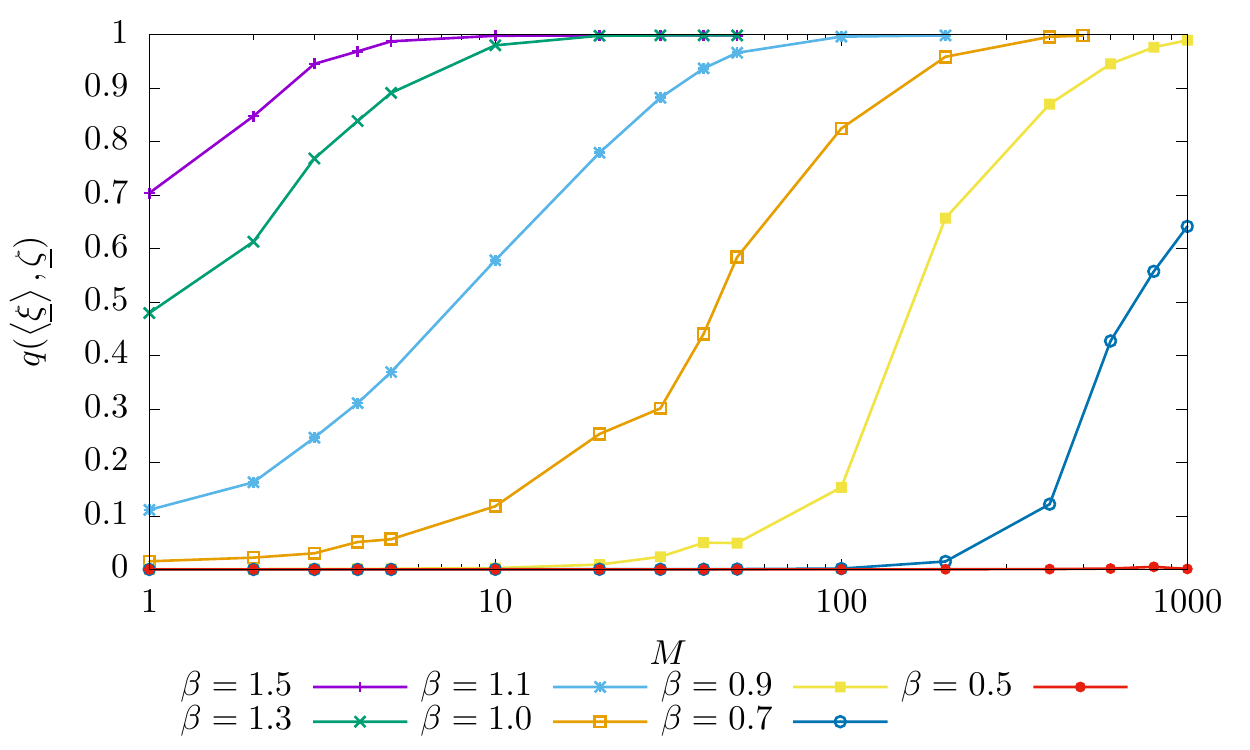}
\caption{Overlap between the teacher's pattern and pattern recovered by the student when using \cref{TAPeq1-upd} with $P=1$, as a function of the number of samples $M$, at different temperatures. System size is $N=1000$. When $\beta<1$, there exists a critical value of $M\sim O(N)$ below which it  is impossible to infer  the pattern, whereas above only a finite set of samples is needed.}
\label{fig:figTAP1}
\end{figure}

For the direct problem, where the coupling matrix is $J_{ij} = N^{-1} \sum_{\mu} \zeta^{\mu}_i \zeta^{\mu}_j $, TAP equations estimate the local magnetizations. For the inverse problem, where we are interested in retrieving the model's parameters given the dataset $\mathcal{D}$, we propose an algorithm based on these equations to estimate the posterior in \cref{eq:dualHMP}. Even if TAP and mean field methods have already been used to solve this inverse problem \cite{braunstein2011inference, cocco2011high,  decelle2016solving, nguyen2012mean, gabrie2015training}, this approach is completely different from the previous ones: dealing with the \textit{dual} model is the key to improve the quality of the reconstructed network. On the \textit{dual} model, the role of spins and patterns is exchanged: $m_i = \left<\xi_i\right>$ and the $M$ sampled configurations play the role of the student's patterns. 
We notice that a similar approach has been indipendently proposed in \cite{hou2019minimal} for an RBM with $2$ hidden binary units, using BP.

We start by considering the simplest case $P=1$.
We introduce a \emph{naive} time indexing for an iterative scheme  of the TAP equations,
\be
m^{t+1}_i  = \tanh\left(\beta \sum_{j=1}^N J_{ij} m^t_j - \frac{\alpha \beta}{1-\beta(1-q^t) } m^t_i \right)\:,
\label{TAPeq1-upd}
\ee
where $J_{ij} = N^{-1} \sum_a s^a_i s^a_i$, $\alpha=P/N$ and $Nq^t=\sum_i (m_i^t)^2$. The entire set of magnetizations $\underline{m}^t$ are updated to achieve $\underline{m}^{t+1}$ in a parallel way.
In principle, any sophisticated time indexing schemes can be employed as long as it achieves the convergence to a physical state.
Particularly, the so-called Approximate Message Passing (AMP) equations has been the focus of many studies in inference problems \cite{zdeborova2016statistical}.
This scheme is inspired by the convergence issues of naive indexing in SK model even in the replica symmetric phase \cite{kabashima2003propagating}.
An explanation to this behavior can be found in \cite{bolthausen2014iterative}, where a less trivial time index setting is shown to improve convergence properties outside the glassy phase.
It was later realized that this time indexing emerges naturally when deriving AMP equations from BP equations, keeping track of BP time indexes. This approach requires considering the fully connected limit of BP equations, that are usually written on sparse graphs \cite{kabashima2001tap,kabashima2003cdma,donoho2009message,krzakala2012probabilistic}.
To our surprise, the AMP equations exhibit convergence issues for the case of Hopfield model in the direct problem, when the initial condition is chosen at random. 
A different approach to derive AMP equations is proposed in \cite{opper2016theory}. This new approach, in our case, does not improve the performances obtained by the \emph{naive} time indexing \cref{TAPeq1-upd}. 
These issues are discussed in detail in \cref{appendix:AMPconvergence}. Thus, in the following, we mainly adopt \cref{TAPeq1-upd}, unless stated otherwise. 

\begin{figure}[h]%
    \centering
    \includegraphics[scale=0.45]{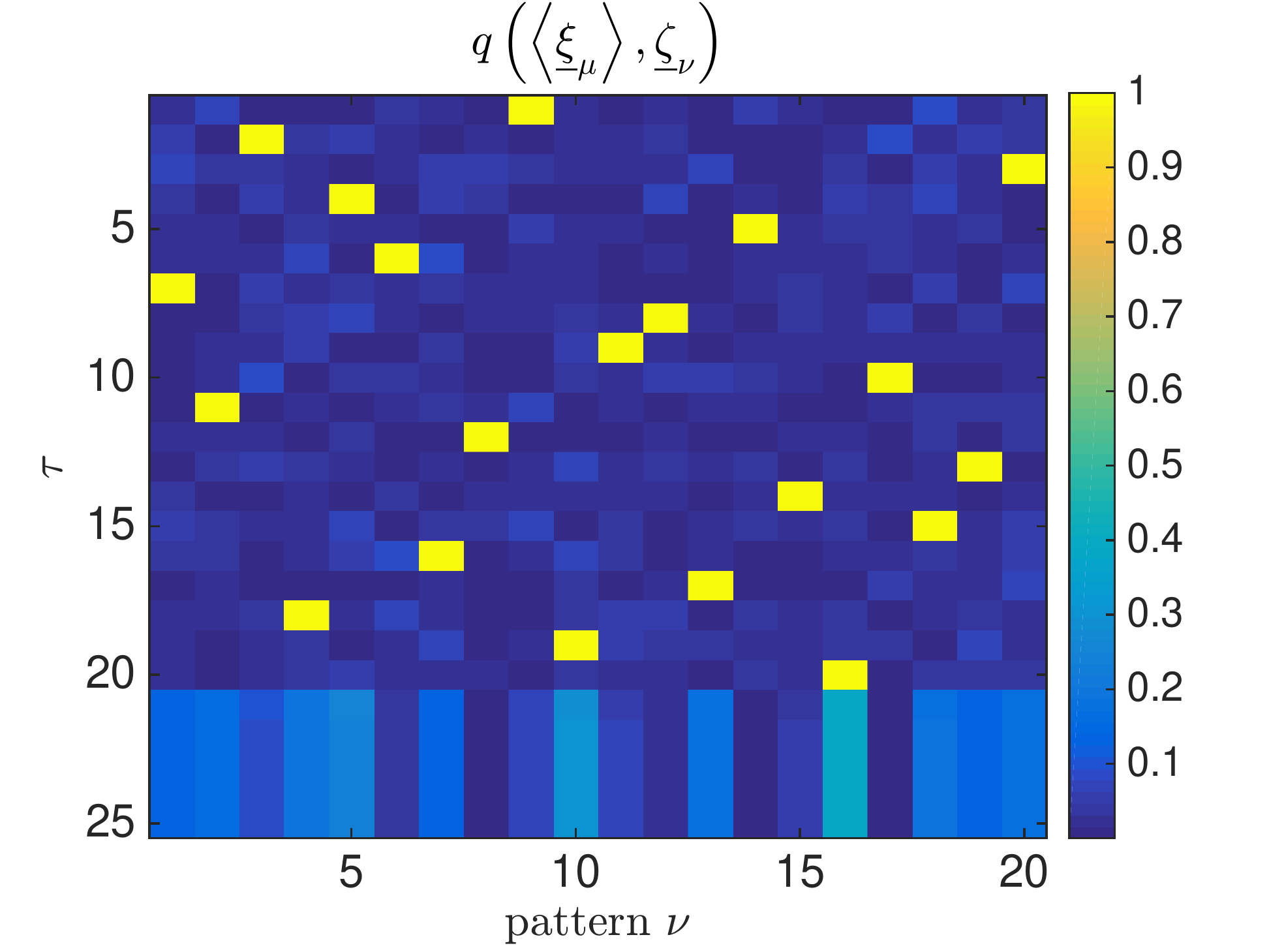}
    \caption{Overlap between the best TAP solutions and the teacher's patterns. The system size is $N=1000$, the teacher generates $P=20$ patterns at $\beta=2$. Inference is done with $P'=25$ students observing only $M=200$ samples, i.e. $10$ per state. At each iteration step $\tau=1\ldots,P'$, we pick the best TAP solution and we plot its overlap with all of the teacher's patterns. We observe clearly that the students are able to retrieve all the patterns from the teacher.}%
    \label{fig:overlap}%
\end{figure}

In Fig. \ref{fig:figTAP1}, we present the teacher-student overlap $q = N^{-1}\sum_i m_i \zeta_i$ for $N=1000$.
We observe that in the ferromagnetic-retrieval phase $\beta>1$,  a perfect reconstruction may be realized already with a small number of samples.
This is due to the large signal contained in the correlation matrix of the data $\overline{c}$. 
In particular, we notice that in the ferromagnetic phase the student is able to find a pattern correlated with the teacher's one even at $M=1$. 
On the other hand in the paramagnetic phase the signal in  $\overline{c}$ is weaker and  reconstruction is possible only exploiting finite size effects, at the price of observing an extensive number of samples. As discussed in \cite{barra2018phase}, the critical fraction $M/N$ of samples necessary for reconstruction corresponds to the paramagnetic-spin glass transition of the direct problem, as long as we restrict the analysis to the Bayes optimal scenario.

 \begin{figure}[b]%
    \centering
    \includegraphics[scale=0.4]{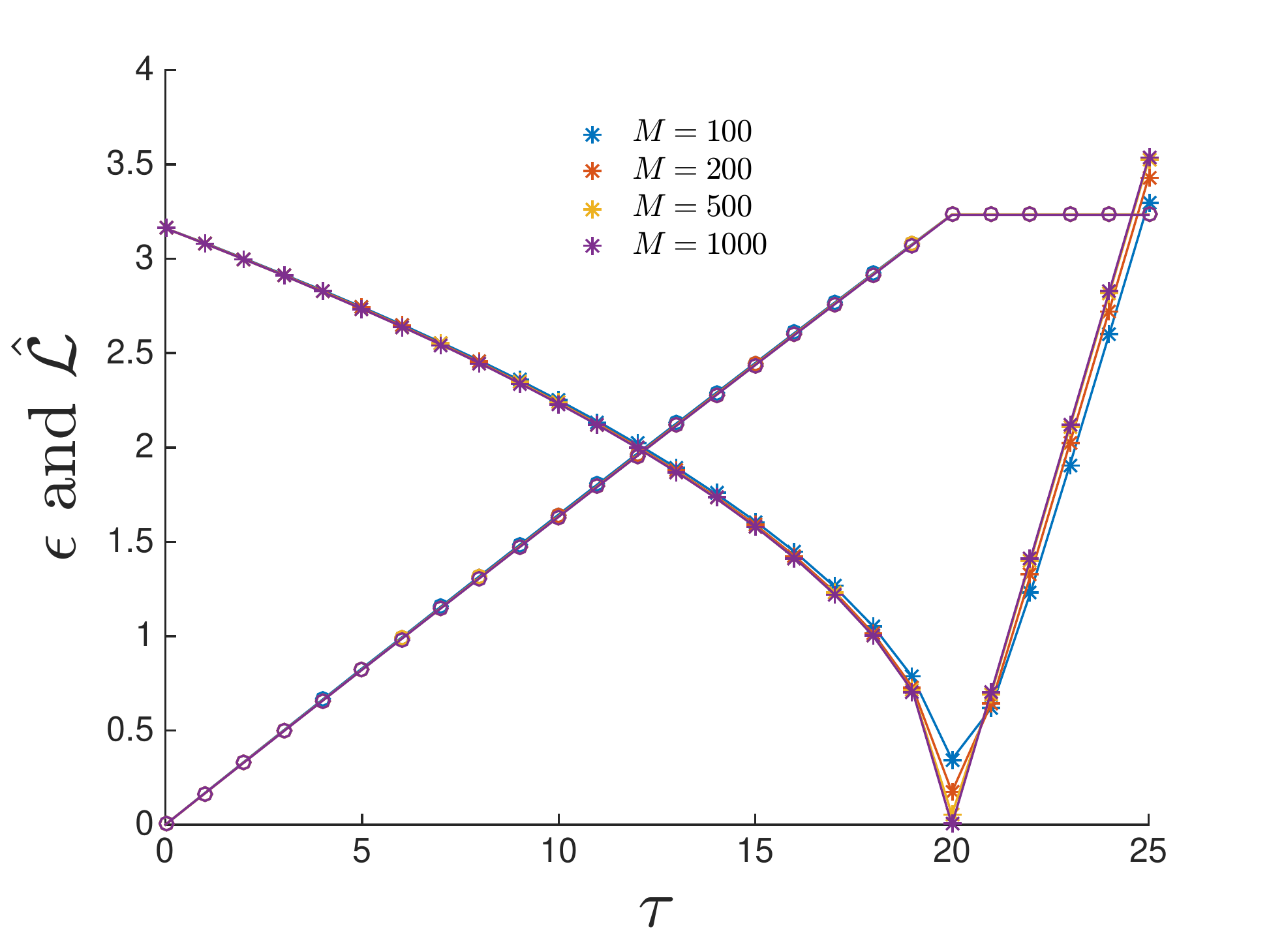}
    \caption{Evolution of the error $\epsilon$ and of the simplified Likelihood $\hat{\mathcal{L}}$, as defined in the text, with the iteration of the algorithm. Different lines refer to different values of $M=100,200,500,1000$ at $\beta=2$, $P=20$, $N=1000$. The error decreases with $M$ and it reaches zero for $M=1000$, although we observe that even with very few samples, the errors are very small and, as shown on Fig. \ref{fig:overlap} the patterns are perfectly recovered. The dependency of $\hat{\mathcal{L}}$ on $M$ is negligible. $\hat{\mathcal{L}}$ is rescaled in order to fit in the figure.}%
    \label{fig:error-M}%
\end{figure}  

 \begin{figure}[b]%
    \centering
    \includegraphics[scale=0.45]{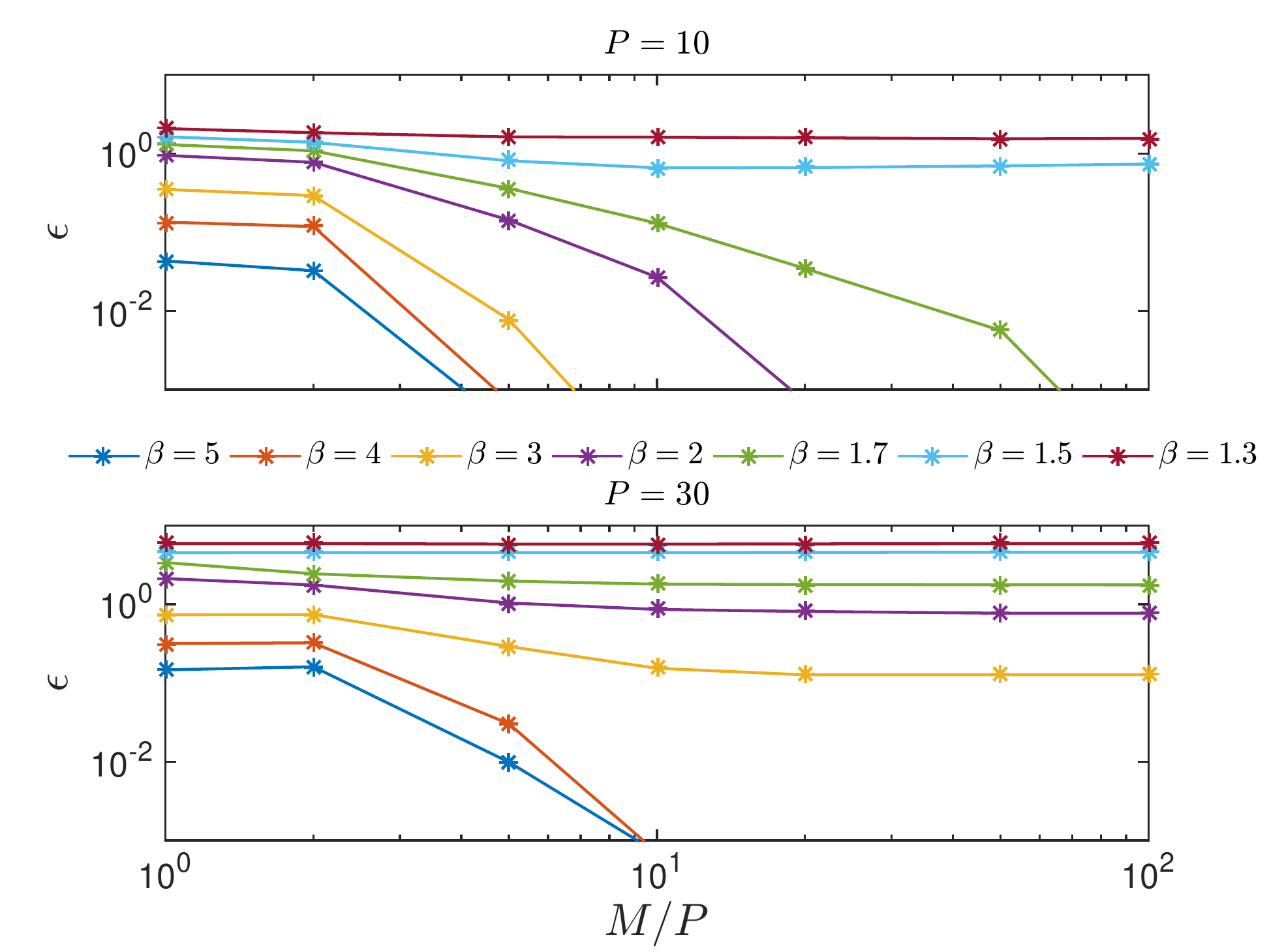}
    \caption{Average error as a function of $M/P$ for a system of size $N=1000$ with a number of patterns $P=10$ (top panel)  and $P=30$ (bottom panel). The reconstruction is done using $P'=2P$ students. The error is computed stopping the algorithm with the criterion described in the text. Each point represent an average over $100$ independent trials. In the retrieval phase, the error goes to zero with $M$.} %
    \label{fig:err-1000}%
\end{figure}

The $P>1$ case is more difficult because of the presence of the denominator in 
\cref{eq:dualHMP}.
However, we argue in \cref{appendix:postP2} that this term is effectively a (soft) orthogonality constraint over inferred patterns. 
This observation allows us to design an inference algorithm accordingly. 
First, let us construct a time evolution of the coupling matrix $J^{\tau}_{ij}$ with its initial condition given by  $J^{\tau=0}_{ij} = N^{-1}  \sum_{a} s_i^a s_j^a  $.
At each time step $\tau$, we consider $P'$ TAP students trying to learn the teacher's patterns \textit{independently}.
Namely, the magnetizations $m_i^{\mu}=\left< \xi_i^{\mu} \right>$ for each student evolve according to \cref{TAPeq1-upd} from a randomly initialized configuration \footnote{To escape the (unstable) fixed point at $m=0$, the absolute value of the local magnetization is chosen to be in the interval $[0.1,1]$.
}.
Upon convergence, we evaluate the $P'$ solutions with the score 
$ S_{\mu} = \sum_{ij}  \sum_{a=1}^M s^a_i s^a_j m^{\mu}_i m^{\mu}_j $.
These scores characterize the quality of the TAP fixed points and we pick the one with the largest score \footnote{This trick is closely reminiscent of the algorithm presented in \cite{franz2019fast}, where the iterative steps are performed by evaluating the likelihood gain obtained moving in different directions and choosing the one with the largest payoff}.
The corresponding magnetization selected by this criterion at time $\tau$ are denoted by $ \underline{m}^{\tau} $. 
Finally, in order to learn the remaining contributions, we remove the rank-1 part associated to the retrieved state $\underline{m}^\tau$ from the coupling matrix.
When the student knows the actual number of patterns, this correspond to the rule $J^{\tau+1}_{ij} = J^{\tau}_{ij}  - \gamma N^{-1} m^\tau_i m^\tau_j$, where $\gamma=M/P$ (assuming that different states are uniformly sampled in the dataset).
We repeat these steps until no further patterns are found.

We stress that our algorithm finds solutions correlated with the patterns without any prior information, i.e. we start iterating the TAP equations from a random initial configuration.
This is a rather remarkable property in comparison to the method used in \cite{braunstein2011inference}, where BP equations were guided to converge to the fixed points associated with the patterns using a reinforcement term aligned with the magnetizations of the states. In Fig. \ref{fig:overlap} we compare the learned $P'$ TAP fix points and the $P$ teacher's patterns in a system of $N=1000$ with $P=20$ at $\beta=2$. In this regime data is generated in the retrieval phase \cite{mezard2017mean}.
The students observe $M=200$ samples, i.e. $10$ samples per state. 
We clearly see that all the $20$ patterns are successfully retrieved from the first 20 students. In addition, let us define two performance measures, the \textit{simplified} likelihood and the reconstruction error
\be
\hat{\mathcal{L}} = \frac{1}{2N}  \sum_{ij} \sum_{\mu=1}^{P'} \sum_{a=1}^M s^a_i s^a_j m^{\mu}_i m^{\mu}_j\:,
\ee
and 
$\epsilon = [N(N-1)/2]^{-1} \sqrt{\sum_{i<j} (J^{r}_{ij}-J^*_{ij})^2 / \sum_{i<j} (J^*_{ij})^2 }$, where $J^*$ denotes the teacher's coupling matrix, and $J^{r}$ is the inferred matrix at time $\tau$, $J^{r}_{ij}=N^{-1}\sum_{t=1}^{\tau}m_i^{t} m_j^{t}$.
The simplified likelihood is defined by neglecting the partition function in 
\cref{eq:dualHMP}.
In \cref{fig:error-M} their behaviors are reported as a function of iteration time. 
As expected, $\epsilon$ decreases as the students learn the patterns but then increases when the students start to learn the remaining noise. 
Similarly, the simplified likelihood $\hat{\mathcal{L}}$ develops a kink at the point where students learn all the patterns, that can be used as a stopping condition of the algorithm.
In Fig. \ref{fig:err-1000} we study the behavior of $\epsilon$ for different values of the temperature. 
As a function of $\beta$, the system sweeps through different regions of the phase diagram.
Data is generated with a sequential Glauber dynamics and states are equally sampled. 
In Fig. \ref{fig:err-1000} we show the behavior of the error computed using the criterion discussed above with the number of observations in different regions of the phase diagram. As expected, perfect reconstruction is obtained only in the retrieval phase. 

\begin{figure}%
    \centering
    \vspace{0.5cm}
    \includegraphics[scale=0.45]{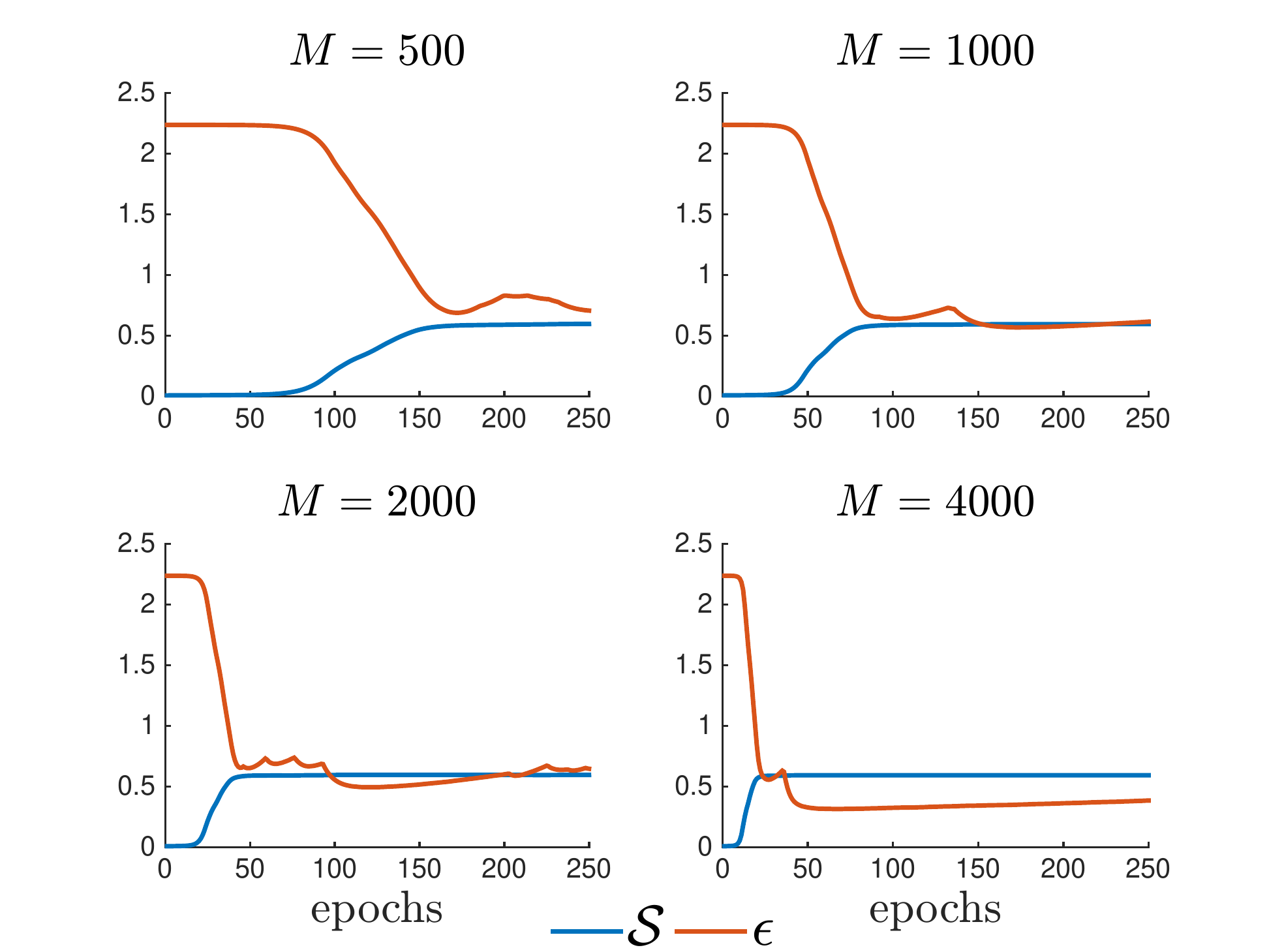}
    \caption{Pseudo-likelihood $\mathcal{S}$ and error $\epsilon$ for $M$ during learning. Data is produced by a teacher Hopfield model with $N=1000$, $P=10$ at $\beta=2$. Learning is done with an RBM with $N_v=N$ visible units and $N_h=15$ hidden units.}%
    \label{fig:learning-curves}%
\end{figure}

\begin{figure}%
    \centering
    \vspace{0.5cm}
    \includegraphics[scale=0.45]{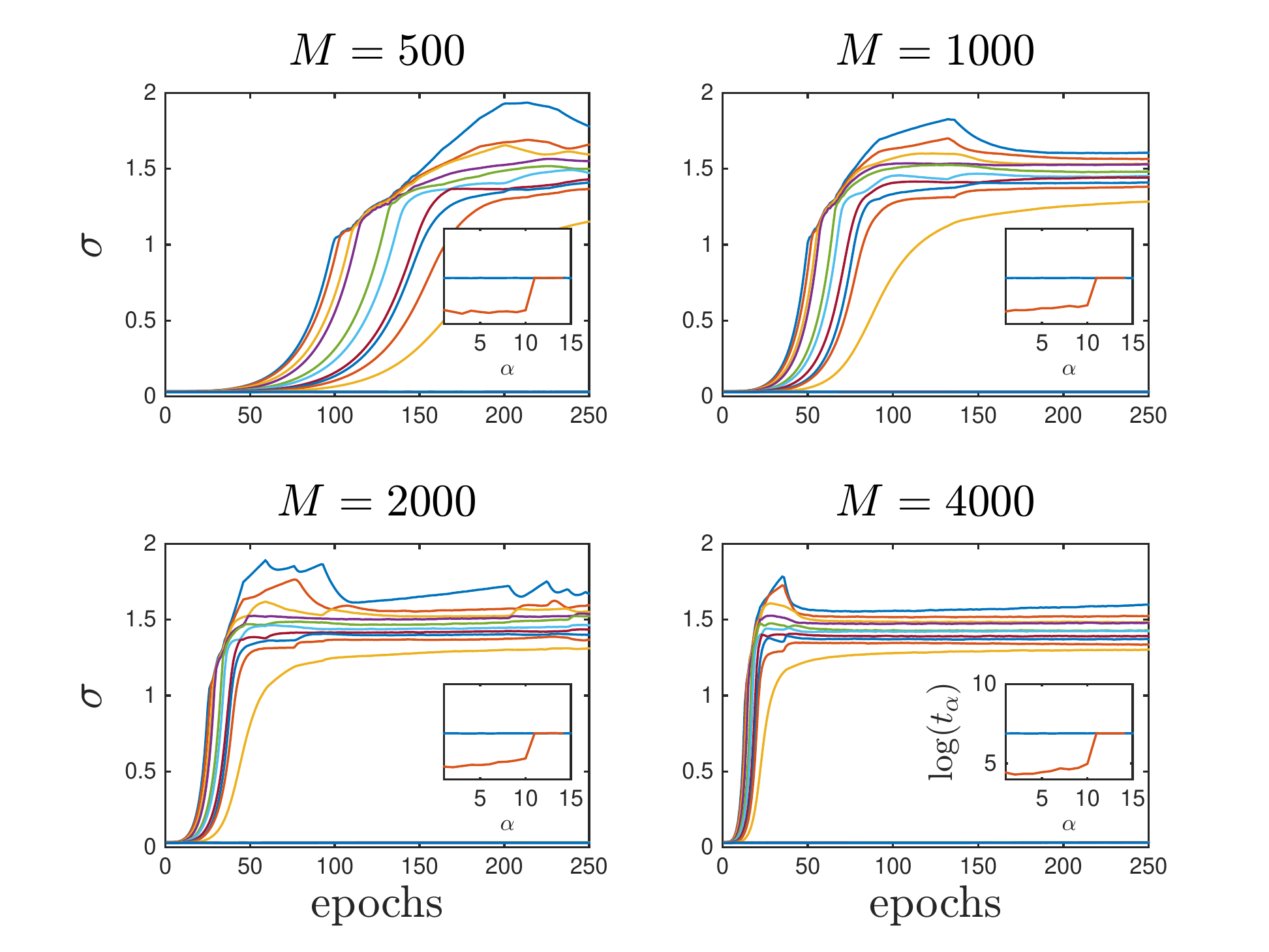}
    \caption{Emergence of singular values $\sigma$ during learning for the same dataset analyzed in Fig. \ref{fig:learning-curves}. $P=10$ modes emerge. Inset: Error $t_{\alpha}$ for different modes at the beginning of learning (blue line) and at the end of learning (orange line). }%
    \label{fig:learning-curves-2}%
\end{figure}

Another method to perform inference is the RBM, which is closely related to the Hopfield model \cite{barra2012equivalence, barra2018phase, mezard2017mean,  agliari2018non, decelle2018thermodynamics}.
In fact, the posterior distribution of 
\cref{eq:dualHMP} can be recast to 
\begin{eqnarray}
&&P(\{W^{\mu}_i\}_{\mu=1,\dots,P}|\mathcal{D}) \propto \nonumber \\
&&Z^{-1}(\underline{W}) \prod_{a=1}^M \int d\underline{\lambda} e^{-\sum_\mu\lambda_{\mu}^2/2 + \sum_{i,\mu} W_i^{\mu} s^a_i  \lambda_{\mu}}\:, \label{eq:RBMdef}
\end{eqnarray}
with $W_i^{\mu} = \sqrt{\frac{\beta}{N}} \xi^{\mu}_i $.
At the same time, 
\cref{eq:RBMdef} defines an RBM with $N_v=N$ binary visible units, and $N_h=P$ Gaussian hidden units. 
Compared to existing methods \cite{sohl2011new,aurell2012inverse,nguyen2012mean,cocco2012adaptive,
decelle2014pseudolikelihood,lokhov2018optimal,franz2019fast}, 
the RBM is both time and space efficient as the number of parameters to be optimized scales as $N_vN_h$ rather than $N^2$.
The number of hidden units $N_h$ plays the role of $P'$. We consider the general setting $N_h \ge P$ in the following. 
Following the standard practice \cite{hinton2012practical}, the weights $W_i^{\mu}$ are learned maximizing the log-likelihood using the PCD-$T$ algorithm, where PCD stands for Persistent Constrastive Divergence \cite{tieleman2008training} and $T$ for the number of Monte Carlo steps used to estimate the part of the log-likelihood derivative involving the partition function. We used $T=10$.
RBM learns a set of weights $J^r_{ij} = \beta^{-1} \sum_{\mu} W_i^{\mu} W_j^{\mu} $ that we can compare with the teacher coupling matrix.
The error between $J^r_{ij}$ and $J^*_{ij}$ decreases during learning but it never achieves the values found with TAP. In order to monitor learning, we study the pseudo-likelihood $\mathcal{S}$ \cite{besag1977efficiency}, i.e a proxy for the likelihood that can be easily computed. In our case, it is defined by $\mathcal{S} =\sum_{r=1}^{N_v}  \mathcal{S}_r$, where
\begin{equation}
\mathcal{S}_r = \frac{1}{M} \sum_{a=1}^{M}  \log( \left<  p(s_r^{a} | \underline{\lambda} \right>_{P(\underline{\lambda}|\underline{s}^a,\{W\} )}  )\:.
\end{equation}
More details on these points as well as on the implementations are given in \cref{appendix:RBM}. In Fig. \ref{fig:learning-curves} we show the behavior of these quantities for the dataset generated by the teacher at $\beta=2$, $N=1000$, $P=10$, when observing a different number $M$ of samples. An RBM with $N_v=1000$ visible units and $N_h=15$ hidden units is used. The minimum of $\epsilon$ is achieved when the pseudo-likelihood flattens. This happen when all of the relevant ($P=10$) modes of the data have been learned, as can be seen in Fig. \ref{fig:learning-curves-2}.
Unless learning starts in the vicinity of the teacher's patterns, final RBM weights do not reproduce them, contrarily to the TAP-based algorithm discussed above. In fact, the Hopfield model is invariant under a rotation in the pattern space \cite{cocco2011high}: the student RBM can learn, at most, the subspace spanned by teacher's patterns. To prove it, we consider the Singular Value Decomposition (SVD) of the dataset, and the SVD of the weights. We denote by $\{\sigma^{\alpha}\}$ the singular values of the matrix $W^{\mu}_i$ and by $t^{\alpha}$ the error in reconstructing the singular vector of the data, indexed by $\alpha$, using only the singular vectors of the weight matrix. In Fig. \ref{fig:learning-curves-2}, we show the emergence of different modes during learning. When the singular values $\sigma^{\alpha}$ of the coupling matrix emerge, the error $t^{\alpha}$ decreases. The first $P=10$ principal modes of the dataset are well represented by the subspace spanned by the singular vectors of the weight matrix $W$.

In summary, we discussed a new method to solve inverse problems with a clusterized dataset. We analyzed the fully connected Hopfield model in a teacher-student scenario and proposed an inference method based on the TAP equations working directly on the posterior distribution, i.e. the \textit{dual} problem.
We discussed a retrieval algorithm based on the parallel updating of the TAP equations with a naive indexing, showing that in our case it gives good results. 
Contrarily to previous methods, our algorithm is able at retrieving patterns, besides couplings, because TAP equations allows to reduces the continuous symmetry under rotation to a simple symmetry under permutation over the pattern labels. 
As a side result, we provide the analysis of the failure of AMP equations in our case, when iterated in a parallel manner starting from a random initial condition.
Finally we compare these results with those obtained with RBM, exploiting their analogies with the Hopfield model. RBM is a good candidate model to perform inference with many variables, a task that would require a much longer execution time to methods based on the optimization of the pseudo-likelihood of an associate pairwise Ising model.
Their ability to perform inference tasks systematically, as well as their performance on inferring sparse models, will be addressed elsewhere. 

We thank S. Franz, F. Ricci Tersenghi and D. Saad for interesting discussions. J.R. and S.H. acknowledge the support of a grant from the Simons Foundation (No. 454941, Silvio Franz). D.T. acknowledges GNFM-Indam  for financial support of the project "Unsupervised Learning with Multi-Layer Boltzmann Machines".

\bibliography{apssamp}
\bibliographystyle{unsrt}

\onecolumngrid

\clearpage
\newpage

\appendix


\section{linear stability analysis of TAP and AMP in the paramagnetic state}
\label{appendix:AMPconvergence}
Here we present the linear stability analysis of TAP and AMP equations in the paramagnetic state. 
We will focus on the direct problem where $J_{ij}$ is constructed from $M$ random patterns. 
While the complete analysis is possible for arbitrary $\alpha$, we find it more instructive to focus on the limit $\alpha \to 0$ as it greatly simplifies the discussion.
As will be shown below, our results are valid if $|1-\beta| \sim O(1)$, which is larger than $O(\alpha)$.

From now on, we denote by TAP the simple iterative updating scheme discussed in the text, reported here for convenience
\be
m^{t+1}_i = \tanh\left(\beta \sum_{j=1}^N J_{ij} m^t_j - \frac{\alpha \beta}{1-\beta(1-q^t) } m^t_i \right)\:
\label{TAPeq1-2}
\ee
and by AMP the iterative scheme derived in \cite{mezard2017mean},
\be
H_i^{t+1} = \frac{1}{1-u^t}\left[ \sum_{j\neq i} J_{ij} m_j^t -u^t H_i^t - \frac{\alpha u^t}{1-u^{t-1}} m_i^{t-1} \right]\:,
\label{AMPeq}
\ee
where $m^t_i=\tanh(\beta H_i^t)$, $u^t=\beta(1-q^t)$ and $Nq^t=\sum_i (m^t_i)^2$. This time index setting naturally emerges from the  expansion of the BP equations in the large connectivity limit \cite{kabashima2001tap}. 

Let us first consider the linear stability of \cref{TAPeq1-2}.
Near the paramagnetic state $M_i \sim 0$, this equation may be expanded into 
\begin{align}
	m^{t+1}_i \simeq \beta \sum_{j=1}^N J_{ij} m^t_j + O(\alpha)
\end{align}
where the second term is neglected as it is of $O(\alpha)$.
Performing the coordinate change with the eigenvectors of $J_{ij}$ as its basis, one obtains 
\begin{align}
	\tilde{m}^{t+1}_\lambda \simeq \beta \lambda \tilde{m}^t_\lambda,
\end{align}
where $\lambda$ is an eigenvalue of $J_{ij}$.
This implies that the paramagnetic solution becomes unstable when $\beta \lambda_{\text{max}} > 1 $.
The spectrum of coupling matrix follows the Marchenko-Pastur law \cite{mezard2017mean}.
Namely, $P$ eigenvalues are $1 + O(\sqrt{\alpha})$ and their eigenvectors span the same space spanned from the set of patterns. The remaining $N-P$ eigenvalues are zero.
Thus we find that the critical temperature is $T_c = 1 + O(\sqrt{\alpha})$ (the true value is $T_c = 1/(1+\sqrt{\alpha})$, found expanding TAP equation beyond the $\alpha\rightarrow 0$ limit, which is identical to the result of replica theory \cite{mezard2017mean}.

\begin{figure}%
    \centering
    \includegraphics[scale=0.4]{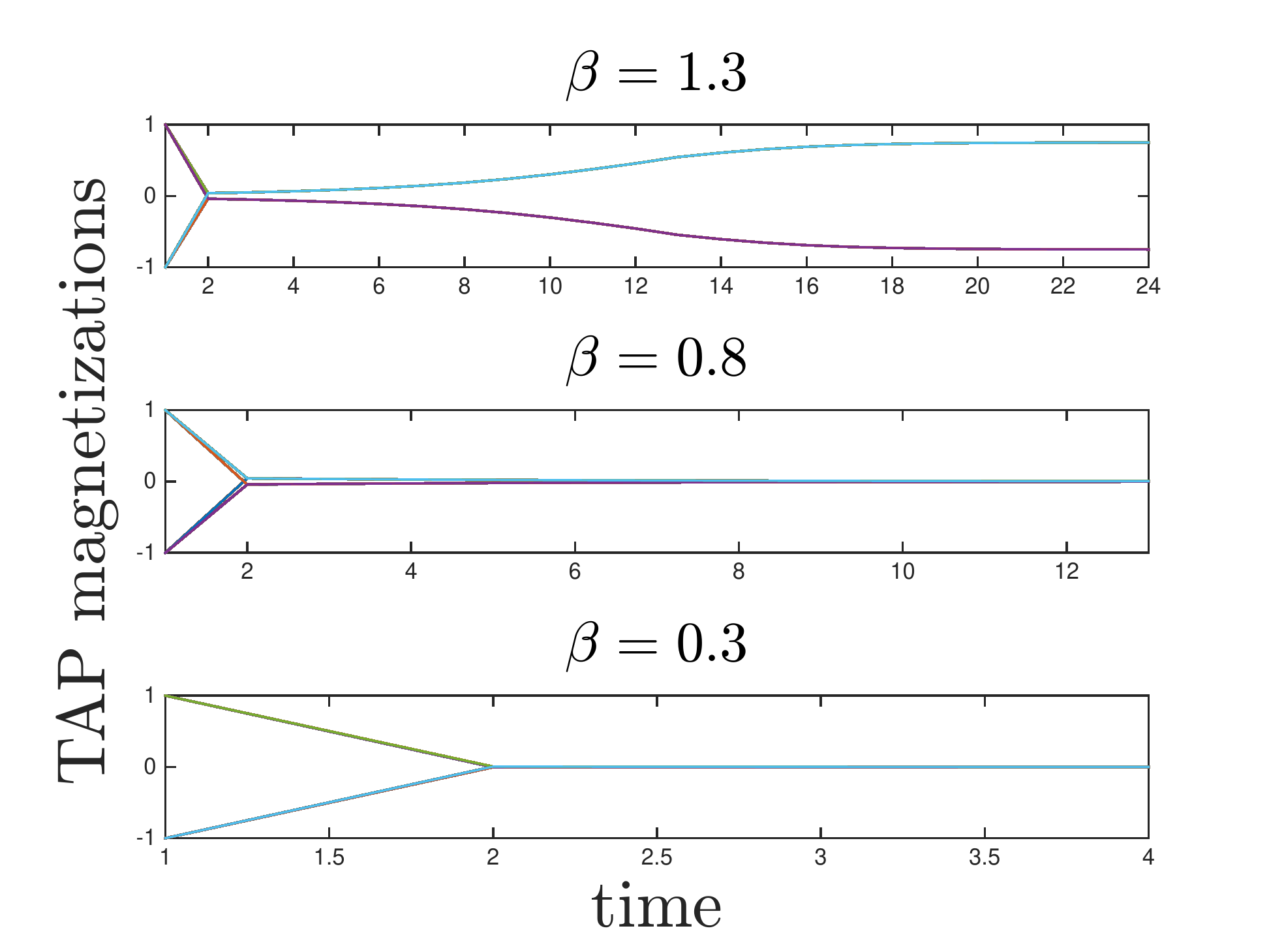}%
    \qquad
    \includegraphics[scale=0.4]{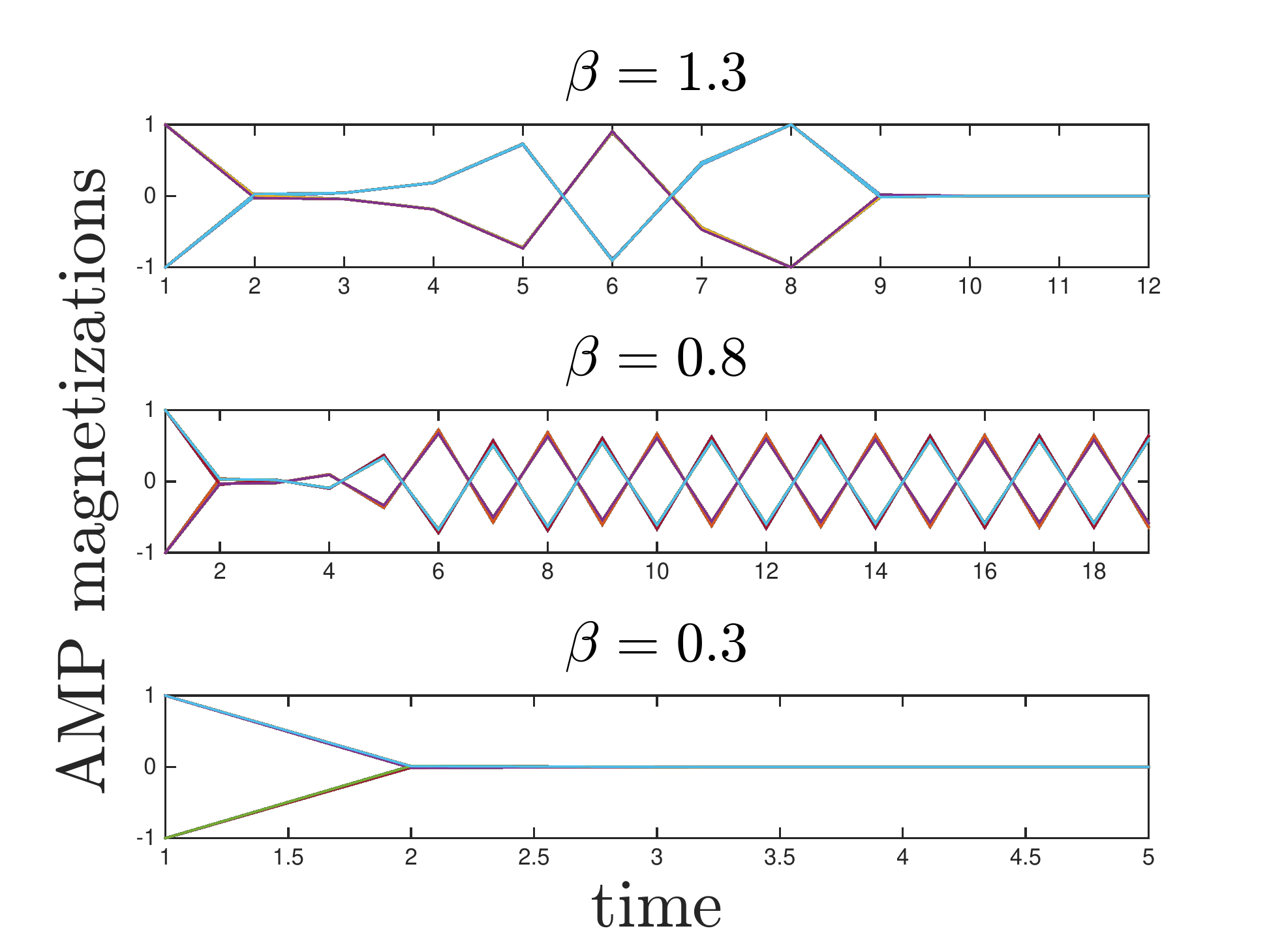}%

    \caption{Trajectories of the $N$ magnetizations $m_i^t$ in the updating schemes of TAP, \cref{TAPeq1-2}, and AMP, \cref{AMPeq}, for three different temperatures at $N=1000$, $P=1$. Most of the trajectories are very similar, thus they are indistinguishable. The critical value is at $\beta=1$. The starting point is chosen at random with the absolute value of the local magnetization equal to one. In the first steps, both TAP and AMP destroy the initial condition and create very small magnetization values. Then, once close to the paramagnetic fixed point $m=0$, AMP eqs. escape from it for $\beta>0.5$ while TAP eqs. do not until $\beta>1$. Moreover, when leaving the paramagnetic state, the direction chosen by AMP is completely random, while TAP moves towards the pattern.}%
    \label{fig:FigTraj}%
\end{figure}

Similarly, AMP \cref{AMPeq} can be expanded as follows:
\begin{align}
\tilde{m}_\lambda^{t+1} = \frac{\beta}{1-\beta}\left[(\lambda -1 )\tilde{m}_\lambda^t - O(\alpha) \right].
\end{align}
Because of the $\lambda-1$ term, in the limit $\alpha\rightarrow 0$, the $N-P$ eigenvalues equal to zero give the largest $O(1)$ contribution to the instability of the paramagnetic fixed point. In particular, the modes associated with patterns, with eigenvalues $1 + O(\sqrt{\alpha})$, give a vanishing contribution. 
From the infinite temperature limit, the first $T$ where this equation becomes unstable is given by $ -\frac{\beta}{1-\beta}  = -1 $, i.e. $T_c = 1/2$. Nevertheless, this unstable direction is orthogonal to the patterns and the magnetization either converges to an unphysical state or never converge (see \cref{fig:FigTraj}). The negative value of the leading eigenvalue for $\beta \in [1/2,1]$ leads to an oscillating behavior starting from the paramagnetic solution, as can be seen in the second plot in \cref{fig:FigTraj}.
Similar issues with parallel updating of the AMP equations were discussed in \cite{zdeborova2016statistical}, and they can be alleviated by updating spins sequentially and introducing a strong dumping. Nevertheless their sequential updating leads to a much slower algorithm, without showing any improvement in the quality of inference in comparison to the parallel updating scheme of \cref{TAPeq1-2}. 

A different updating scheme of the TAP equations has been recently proposed by \cite{opper2016theory}. This approach does not require to consider the fully connected limit of the BP equations and it is suitable to be applied in systems with dense random coupling matrices. It is based on a dynamical mean field theory which allows to study the dynamics of iterative algorithms in the thermodynamic limit by averaging over the noise contained in the couplings. For the Hopfield model, the updating scheme turns out to be 
\begin{align}
m_i^{t+1} &= \tanh \left( z_i^t + A_t m_i^t\right)\\
z_i^t &= A_t \left( \sum_j J_{ij} m_j^t -m_i^t \right) + \alpha (1-q^t) A_t z_i^{t-1}
\label{TAP-OW}
\end{align}
where $A_t = \beta / (1+\alpha u_t)$ and $u_t$ is the same quantity introduced in the AMP \cref{AMPeq}.
It is possible to see that this updating scheme does not present the issues of the AMP algorithm by repeating the same $\alpha \rightarrow 0$ analysis presented above.

\begin{figure}%
    \centering
    \vspace{0.5cm}
    \includegraphics[scale=0.6]{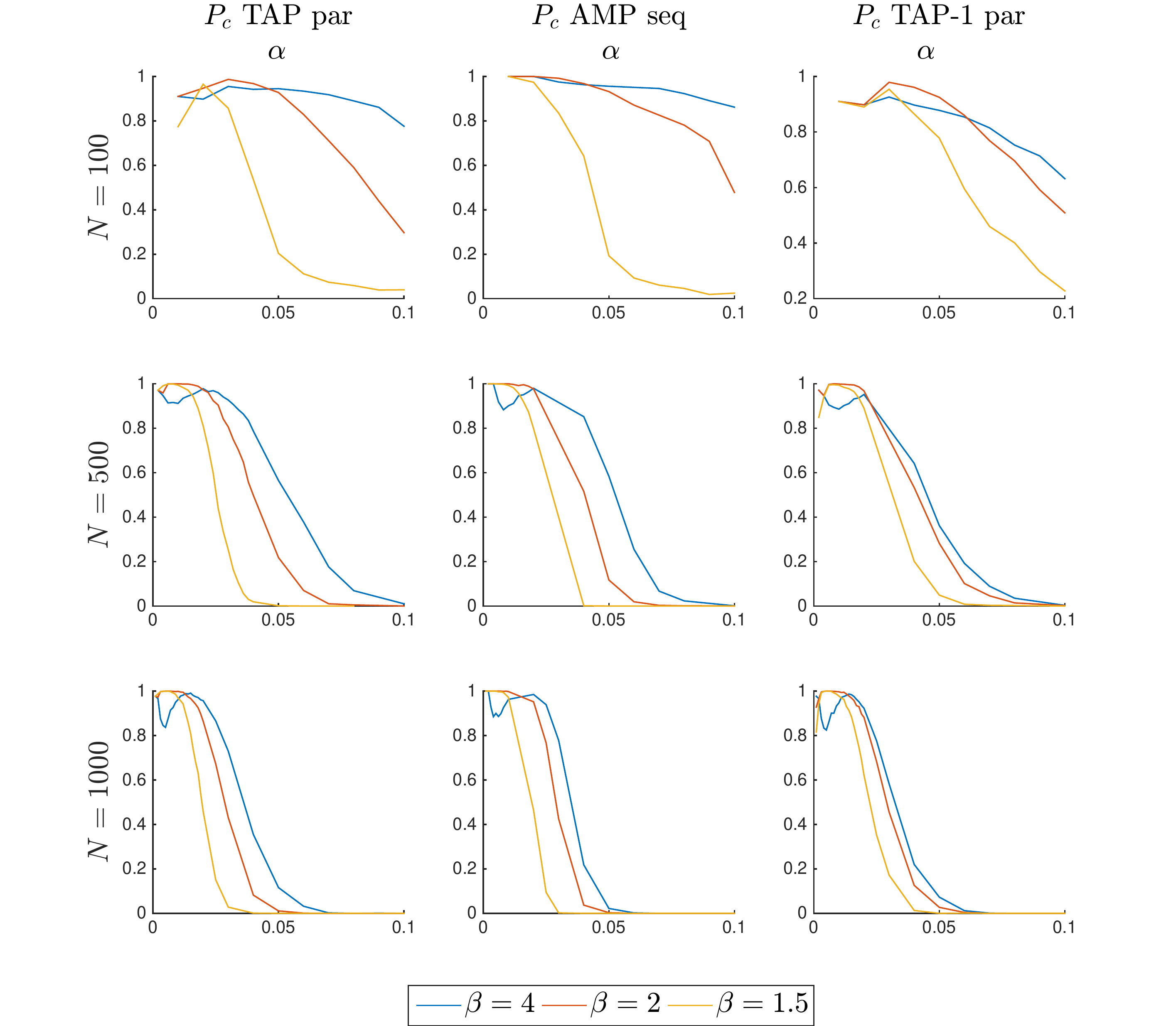}
    \caption{Probability to converge to a pattern when iterating \cref{TAPeq1-2} (left), \cref{AMPeq} (center), \cref{TAP-OW} (right), starting from a random initial condition in the direct problem. This probability is estimated running $1000$ independent experiments from different realizations of the patterns and different initial conditions and counting the number of times that the equations converged to one of the patterns of the system with overlap greater than $0.7$, in order to exclude mixture states. The sequential updating of the AMP equations is done with a dumping term equal to $0.95$. The performance of all these algorithms is similar, with the second one being much slower. The initial absolute value of the local magnetizations are mostly irrelevant in the first two cases (and it is chosen to be $1$), but needs to be chosen small at low temperatures in the third case (and it is chosen to be $0.1$).}%
    \label{fig:conv-analysis}%
\end{figure}

In Fig. \ref{fig:conv-analysis} we compare the performances of these three algorithms for different system sizes. We define $P_c$ as the probability to converge to one of the patterns of the system with overlap greater than $0.7$ when the initial condition is chosen at random. Sequential AMP were iterated with a damping term $d$, i.e. $m_i^{t+1} = (1-d) \tanh \beta H_i^{t+1} + d m_i^t$, and $d=0.95$. For the two parallel TAP equations, (\cref{TAPeq1-2}-\cref{TAP-OW}), the iteration is stopped when the average difference between $m_i^{t+1}$ and $m_i^{t}$ is smaller than $0.001$. For the AMP sequential algorithm the iteration is stopped when the average between  $\tanh \beta H_i^{t+1}$ and $\tanh \beta H_i^{t}$ is smaller than $0.001$. In all the cases we observe that convergence to patterns is achieved in the retrieval phase. For small values of $N$, due to finite size effect, convergence regime extends in the metastable retrieval phase too. 

The instability issue of the AMP equations presented above holds for the direct problem, but it can be extended also to the inverse, \textit{dual} problem. In this last case, where $J_{ij}=N^{-1}\sum_{a=1}^M s^a_i s^a_j$, if there is enough signal in the data and $\lambda_{\text{max}} > 2$, inference is possible also with parallel AMP equations. 
Nevertheless the analysis shows that obtaining time indexes from BP does not necessarily lead to good algorithms.  
TAP, as well as BP, equations describe only fixed points of the associated free energy and, in principle, any updating scheme could be used to solve these equations in an iterative manner, as shown in \cite{opper2016theory}.
The relevance of this observation for other problems requires further analysis and, given that the AMP convergence issues are usually mitigated by considering a sequential updating with a strong dumping, it would be interesting to study whether a similar improvements is achieved when iterating TAP equations with the \textit{naive} time indexing sequentially and with a strong damping, in problems where their parallel updating was failing.

\section{Posterior for $P>1$}
\label{appendix:postP2}

We discuss the role of the difficult term arising in the posterior distribution when $P>1$. We show that for the simple case $P=2$, it has  a clear interpretation in terms of a constraint on the orthogonality of the inferred patterns. In fact, let us consider 
\be
Z(\{\xi^1_i,\xi^2_i\}) = \sum_{s} e^{\frac{\beta}{N}\sum_{ij}(\xi^1_i \xi^1_j+\xi^2_i \xi^2_j) s_i s_j}\:,
\label{eqZ12}
\ee
and let us define $S=\{i : \xi^1_i = \xi^2_i \}$, such that $|S|= N(1+q)/2$, where $q$ is the mutual overlap between the two patterns, $Nq = \sum_i \xi^1_i \xi^2_i$. The exponent in \cref{eqZ12} reads 
\be
H_{\xi}(s) = \frac{ 2 \beta}{N}\sum_{i\in S, j \in S}\xi^1_i \xi^1_j s_i s_j + \frac{ 2 \beta}{N}\sum_{i\in \overline{S}, j \in \overline{S}}\xi^1_i \xi^1_j s_i s_j \:,
\label{Hxidec}
\ee
where we indicate with $\overline{S}$ the complement of set $S$. Using again the gauge transformation $s'_i=\xi^1_i s_i$, \cref{Hxidec} leads to 
\be
Z(\{\xi^1_i,\xi^2_i\}) = Z_{cw,\beta (1+q)}^{N(1+q)/2} Z_{cw,\beta (1-q)}^{N(1-q)/2}\:,
\ee
where we indicate with $Z_{cw,\beta}^N$ the partition function of a ferromagnetic Curie-Weiss model at inverse temperature $\beta$.
We observe that the interaction depends only on their mutual overlap. If we define $\phi = -N \log Z$, we obtain
\be
\phi(q) = \frac{1+q}{2}f_{cw}(\beta (1+q)) + \frac{1-q}{2}f_{cw}(\beta (1-q))
\ee
where $f_{cw}(\beta)$ is the free energy of the Curie-Weiss model at inverse temperature $\beta$. It is easy to check that $\phi(q)$ is a convex function with a minimum in $q = 0$. Thus the term $- M\log Z(\underline{\xi})$ in the posterior can be interpreted as a soft regularizer for patterns orthogonality.

\section{Restricted Boltzmann Machine}
\label{appendix:RBM}

A Restricted Boltzmann Machine (RBM) is a particular kind of Boltzmann machines in which units are divided in two layers, formed by visible $\{s_i\}$ and hidden $\{\lambda_{\mu}\}$ units, and only interactions $W^{\mu}_i$ between units of different layers are allowed, such that the proxy probability distribution reads
\begin{equation}
P(\underline{s},\underline{\lambda}|\{W\})=Z^{-1}(\{W\}) e^{-E_W(\underline{s},\underline{\lambda})}
\end{equation}
where $E_W(\underline{s},\underline{\lambda}) = - \sum_{i,\mu} W^{\mu}_i s_i \lambda_{\mu}$, and $Z(\{W\}) $ is the partition function, \begin{equation}
Z(\{W\}) = \int \prod_{\mu=1}^{N_h} d P(\lambda_{\mu}) \sum_{\underline{v}} e^{-E_W(\underline{s},\underline{\lambda})}\:.
\end{equation}
For our purposes, $P(\underline{\lambda})$ denotes a generic distribution over hidden units, while visible units are $\pm 1$ binary variables. We indicate with $N_v$ the number of visible units and with $N_h$ the number of hidden units. 
RBM has the property that the two conditional probabilities, $P(\underline{s}|\underline{\lambda},\{W\})$ and $P(\underline{\lambda}|\underline{s},\{W\})$, factorize over the visible (resp. hidden) units.
These machines are used to learn weights such that the distribution over the visible units reproduce the distribution of the data. In other words  
\begin{equation}
P(\underline{s}|\{W\}) = \int \prod_{\mu=1}^{N_h} d P(\lambda_{\mu})  P(\underline{s},\underline{\lambda}|\{W\})
\end{equation}
should reproduce as close as possible $P_{D}(\underline{s}) = M^{-1} \sum_{a=1}^M \delta_{\underline{s},\underline{s}^{a}}\:. $
Weights can be found  minimizing the KL distance between the two distribution, which is equivalent to maximizing the likelihood $\prod_{a=1}^M P(\underline{s}^a|\{W\}) $ or the log-likelihood
\begin{equation}
\mathcal{L}=\frac{1}{M} \sum_{a=1}^M\left( -\log Z(\underline{W})+ \log \int  d P(\underline{\lambda})   e^{-E_W(\underline{s}^a,\underline{\lambda})} \right) \:.
\end{equation}
Optimal weights can be learned by gradient ascent:
\begin{equation}
W^{\mu}_i = W^{\mu}_i + \left( \left< \lambda_{\mu} s_i\right>_{D} - \left< \lambda_{\mu} s_i\right>_{RBM}  \right)
\label{sgdpone}
\end{equation} 
where the first average, usually referred to as positive phase, is 
\begin{equation}
\left< \lambda_{\mu} s_i \right>_{D} = M^{-1} \sum_a^{M}  \int  d P(\underline{\lambda}) P(\underline{\lambda}|\underline{s}^a,\{W\}) \lambda_{\mu} s_i^a
\end{equation}
and the second average, usually referred to as negative phase, is
\begin{equation}
\left<  \lambda_{\mu} s_i \right>_{RBM} = \frac{\partial}{\partial W_i^{\mu}} \log Z(\underline{W})\:.
\end{equation}
The second one is known to be difficult and it can be computed with approximate methods. One way to estimate it is to use a Monte Carlo (MC). Depending on the number of steps $T$ of the MC Markov chain, this method is referred to as CD-$T$, where CD stands for Contrastive Divergence. In the text, we discuss results obtained with $T=10$. 
When the positive term is computed over a sub set (mini-batch) of the dataset, the direction indicated by the gradient does not correspond to the correct one obtained considering the whole dataset. This trick introduces a source of randomness in the path to the solution, and the associate learning algorithm is called Stochastic Gradient ascent. In our experiments we use a mini-batch size equal to $100$. 
Since mini-batch samples are independent, different parallel MC can be used. In our experiments we used $100$ MC chains, one per mini-batch sample. Their initial conditions can be chosen to be the considered samples, but this quickly results in over fitting the parameters, since the MC dynamics spend all the time in the phase space regions close to the samples.
When the initial condition of the MC dynamics is chosen at random and we keep track of their positions through different batches and epochs, this method is called Persistent CD (PCD). Our results are obtained using PCD.

As stated above, the likelihood function cannot be easily computed. Thus, we introduce the pseudo-likelihood that, for a model with hidden units, is defined by $\mathcal{S} =\sum_{r=1}^{N_v}  \mathcal{S}_r$, where
\begin{equation}
\mathcal{S}_r = \frac{1}{M} \sum_{a=1}^{M}  \log( \left<  p(s_r^{a} | \underline{\lambda} \right>_{P(\underline{\lambda}|\underline{s}^a,\{W\} )}  )\:,
\end{equation}
where the term inside the $\log$ is defined by
\begin{equation}
\left<  p(s_r^{a} | \underline{\lambda} \right>_{P(\underline{\lambda}|\underline{s}^a,\{W\} )} =  \int  d P(\underline{\lambda}) p(s_r^{a} | \underline{\lambda} ) P(\underline{\lambda}|\underline{s}^a,\{W\}) 
\label{eq:probilog}
\end{equation} 
and it is equal to 
\begin{equation}
\left<  p(s_r^{a} | \underline{\lambda} \right>_{P(\underline{\lambda}|\underline{s}^a,\{W\} )} = \mathcal{N}^{-1}_a \int  d P(\underline{\lambda}) e^{ \sum_{\mu k} W_k^{\mu} s_k^a \lambda_{\mu}}\:,
\end{equation}
where $\mathcal{N}_a$ is a sample dependent normalization factor, 
\begin{equation}
\mathcal{N}_a = \sum_{s_r^a} \int  d P(\underline{\lambda}) e^{ \sum_{k \mu} W_k^{\mu} s_k^a \lambda_{\mu}}.
\end{equation}
The Pseudo-likelihood is optimized by the same set of parameters $\{W\}$ that optimize the likelihood. In order to show this property, we can take derivatives of $\mathcal{S}$:
\begin{dmath}
\frac{\partial \mathcal{S}_r}{\partial W_{r}^{\nu}} = \frac{1}{M} \sum_{a=1}^M \left( \frac{ \int \prod_{\mu} d P(\lambda_{\mu})s_r^a \lambda_{\nu}  e^{\sum_{k \mu} W_k^{\mu} s_k^a  \lambda_{\mu}} }{ \int d P(\underline{\lambda}) e^{ \sum_{k \mu} W_k^{\mu} s_k^a \lambda_{\mu}} }   -  \frac{\partial}{\partial W_r^{\nu}} \log \sum_{s_r^a} \int d P(\underline{\lambda}) e^{ \sum_{k \mu} W_k^{\mu} s_k^a \lambda_{\mu}}\right) \:.
\label{pseudolikeder}
\end{dmath}
The definition 
\begin{dmath}
P(\underline{\lambda}|\underline{s}^a,\{W\} )  =  \frac{   e^{\sum_{k,\mu} W_k^{\mu} \lambda_{\mu} s_k^a}  }{ \int d P(\underline{\lambda}) e^{ \sum_{k, \mu} W_k^{\mu} \lambda_{\mu}s_k^a } } 
\end{dmath}
allows to write the first term of \cref{pseudolikeder} as 
\begin{equation}
\left< \lambda_{\nu} s_r \right>_{D} = \frac{1}{M} \sum_{a=1}^M \int d P(\underline{\lambda}) \lambda_{\nu} s^a_r P(\underline{\lambda}|\underline{s}^a,\{W\})\:.
\end{equation}
The second term is given by the average over samples of
\begin{dmath}
\mathcal{N}_a^{-1} \frac{\partial \mathcal{N}_a}{\partial W^{\nu}_r} = \frac{\sum_{s_r^a} \int d P(\underline{\lambda}) s_r^a \lambda_{\nu} e^{ \sum_{k \mu} W_k^{\mu} s_k^a \lambda_{\mu}}}{\sum_{s_r^a} \int d P(\underline{\lambda}) e^{ \sum_{k \mu} W_k^{\mu} s_k^a \lambda_{\mu}} }
\end{dmath}
and similar manipulations on the second term lead to
\begin{dmath}
\frac{\partial \mathcal{S}}{\partial W_{r}^{\nu}}  = \frac{1}{M} \sum_{a=1}^M s_r^a  \langle \lambda_{\nu}  \rangle_{P(\underline{\lambda}|\underline{s}^a,\{W\})}
 -   \frac{1}{M} \sum_{a=1}^M \langle \lambda_{\nu} \tanh\sum_{\mu} W^{\mu}_r \lambda_{\mu} \rangle_{P(\underline{\lambda}|\underline{s}^a,\{W\})}  = \langle     \lambda_{\nu} s_r \rangle_{D} - \langle  \lambda_{\nu} \tanh \sum_{\mu} W^{\mu}_r \lambda_{\mu}  \rangle_{D}\:.
\end{dmath}
In the infinite sampling limit, 
\begin{dmath}
\lim_{M\rightarrow \infty} \langle  \lambda_{\nu} \tanh \sum_{\mu} W^{\mu}_r \lambda_{\mu}  \rangle_{D}  = \langle  \lambda_{\nu} s_r  \rangle_{RBM} \:.
\end{dmath}
In fact, it is easy to show that 
\begin{dmath}
\langle  \lambda_{\nu} \tanh \sum_{\mu} W^{\mu}_r \lambda_{\mu}  \rangle_{RBM}  = \langle  \lambda_{\nu} s_r  \rangle_{RBM} \:
\end{dmath}
and, on the other hand, that 
\begin{equation}
 \lim_{M\rightarrow \infty} \langle  \lambda_{\nu} \tanh \sum_{\mu} W^{\mu}_r \lambda_{\mu}  \rangle_{D}  = \langle  \lambda_{\nu} \tanh \sum_{\mu} W^{\mu}_r \lambda_{\mu}  \rangle_{RBM} \:.
\end{equation} 
Thus the gradient of the pseudo-likelihood $\mathcal{S}$ vanishes on the same set of parameters $\{W\}$ that solve $0=\partial_{W} \mathcal{L}$, and this is the reason the pseudo-likelihood can be used to control the learning state.
In practice, the probability in \cref{eq:probilog}, can be estimated after one step of Monte Carlo:

\begin{equation}
\left<p(\underline{s_r}^a|\underline{\lambda})\right>_{P(\underline{\lambda}|\underline{s}^a,\{W\})} \sim \frac{e^{s_r^a \sum_{\mu} W^{\mu}_i \lambda_{\mu}}}{2 \cosh \sum_{\mu}W^{\mu}_i \lambda_{\mu} }
\end{equation}
where $\underline{\lambda}$ is sampled from the distribution $P(\underline{\lambda}|\underline{s}^a,\{W\})$.

Finally, we discuss the learning of the RBM compared to our TAP based algorithm. As mentioned in the text, unless learning starts in the vicinity of the teacher's patterns, final RBM weights do not reproduce them. In fact, the Hopfield model is invariant under a rotation in the pattern space. The dataset $\mathcal{D}$ analyzed by the student could have been produced by another set of patterns $\{\underline{\hat{\zeta}}^{\mu}\}_{\mu=1,\dots,P}$ given by $\hat{\zeta}^{\mu}_i = \sum_{\nu=1}^P O_{\mu,\nu}\zeta^{\nu}_i$ where $O$ is an orthogonal matrix. This symmetry implies that the student RBM cannot learn exactly the teacher's patterns. One could think that the singular vectors of $W$ should learn at least the principal vectors of the data (that, given the spherical symmetry, are not necessarily aligned along the teacher's patterns), as discussed in \cite{decelle2018thermodynamics}. Nevertheless this is true only during the initial steps of learning, when couplings are small. 
This is reminiscent of the results discussed in \cite{cocco2011high}, where the posterior of the problem is analyzed in a perturbative expansion. 
At the first order, corresponding to the small couplings regime, the student's patterns are aligned along the singular vectors of the data at zero order. Anyway, computing higher order corrections, this relation breaks down.  

In the following we show that RBM is learning the subspace spanned by the singular vectors of the data. To prove it, we consider the Singular Value Decomposition SVD of the dataset, $D=U_{D} \Sigma_{D} V_{D}^T$, where, considering $N<M$, $D$ is a $N \times M$ matrix, $U_D$ is a orthogonal $N \times N$ matrix, $\Sigma_D$ is a $N \times M$ matrix, with only $N$ diagonal elements different from zero, and $V_D$ is a orthogonal $M \times M$ matrix. $D$ represent the matrix of the dataset $\mathcal{D}$, where each column is a sample. Similarly, we consider the SVD of the weight matrix, $W^T=U_{W} \Sigma_{W} V_{W}^T$, where $W^T$ is $N_v \times N_h$, $U_{W}$ is a $N_v \times N_v$ orthogonal matrix, $\Sigma_{W}$ is a $N_v \times N_h$ matrix, with $N_h$ diagonal elements different from zero, and $V_{W}$ is a $N_h \times N_h$ orthogonal matrix. We consider $N_v=N$ and we decompose all of the data modes $u^{(\alpha),D}_i= [U_{D}]_{i \alpha}$ onto the subspace spanned by the first $N_h$ singular vectors of the weights, $u^{(\mu),W}_i= [U_{W}]_{i \mu}$, $\mu=1,\ldots,N_h$:
\begin{equation}
\vec{u}^{(\alpha),D}=\sum_{\mu=1}^{N_h} c_{\mu}^{\alpha} \vec{u}^{(\mu),W} + e^{\alpha}(W) \\,\quad c_{\mu}^{\alpha}=\left<\vec{u}^{(\mu),W},\vec{u}^{(\alpha),D}\right>
\end{equation}
where $\{ \vec{u}^{(\mu),W} \}_{\mu=1,\ldots,N_h}$ are orthogonal vectors normalized to one. We measure the behavior of 
\begin{equation}
t_{\alpha}=\sum_i|e^{\alpha}_i|/\sum_i|u^{(\alpha),D}_i|
\label{deferrt}
\end{equation}
at the beginning and at the end of learning. This quantity measures the difference between the original vector and its projection onto the subspace spanned by the basis $\{ \vec{u}^{(\mu),W} \}_{\mu=1,\dots N_h}$.
The results of this analysis are found in the insets of Fig. \ref{fig:learning-curves-2}, where we plot these quantities at the initial stage of learning and at the end. 

\end{document}